\documentclass[lettersize,journal]{IEEEtran}
\usepackage{amsmath,amsfonts}
\usepackage{algorithm}
\usepackage{algorithmic}
\usepackage{array}
\usepackage[caption=false,font=normalsize,labelfont=sf,textfont=sf]{subfig}
\usepackage{textcomp}
\usepackage{stfloats}
\usepackage{url}
\usepackage{verbatim}
\usepackage{graphicx}
\usepackage{cite}
\usepackage{diagbox}
\usepackage{amsmath}
\usepackage{multirow}
\usepackage{makecell}
\usepackage{setspace}
\usepackage{threeparttable}
\usepackage{color}
\usepackage{amsthm}

\usepackage{booktabs}
\usepackage{graphicx}

\def\BibTeX{{\rm B\kern-.05em{\sc i\kern-.025em b}\kern-.08em
    T\kern-.1667em\lower.7ex\hbox{E}\kern-.125emX}}
\usepackage{balance}
\begin{document}
\title{\LARGE
Quantifying Cyber-Vulnerability in Power Electronics Systems via an Impedance-Based Attack Reachable Domain
}

\author{
Hongwei Zhen,~\IEEEmembership{Student Member,~IEEE,}
Ze Yu,~\IEEEmembership{Student Member,~IEEE,}
Xin Xiang,~\IEEEmembership{Member,~IEEE,}\\
Wuhua Li,~\IEEEmembership{Senior Member,~IEEE,}
Mingyang Sun,~\IEEEmembership{Senior Member,~IEEE}
\vspace{-30pt}

}

\maketitle

\begin{spacing}{0.9}
\begin{abstract}
Power electronics systems are increasingly exposed to cyber threats due to their integration with digital controllers and communication networks. However, an attacker-oriented metric is still lacking to quantify the extent to which a node can be pushed toward instability within a privilege-constrained action space. This letter proposes an impedance-based Attack Reachable Domain (ARD) framework that maps feasible adversarial actions to critical-eigenvalue migration through impedance reshaping. Based on the ARD, an Attack Penetration Index is defined to quantify node-level cyber-vulnerability by jointly characterizing the penetration of the nominal stability margin and the accessibility of successful destabilizing attacks within a privilege-constrained action space. To make the proposed assessment computable when inverter models are unavailable, a practical gray-box workflow is further established by integrating existing impedance identification and differentiable surrogate tools. Case studies on a 4-bus system and a modified IEEE 39-bus system show that coordinated cross-layer manipulations are markedly more damaging than isolated single-layer attacks, and that the proposed metric reveals vulnerability patterns that cannot be inferred from grid-strength indicators.
\end{abstract}


\vspace{-12pt}

\section{Introduction}
\IEEEPARstart{P}{ower} electronic systems (PESs) are increasingly integrated with digital controllers, communication links, and software-configurable control stacks, expanding their attack surface and making cybersecurity directly relevant to dynamic stability~\cite{Sahoo2021multi}.
Recent incidents and vulnerability reports on solar plants, wind generation assets, and smart inverters~\cite{johnson2025public} further indicate that cyber attacks are no longer a peripheral concern, but a stability relevant risk for converter-dominated grids.

Existing cybersecurity research on PESs has advanced along several important directions. Prior studies have investigated exposed attack surfaces of inverter-interfaced devices\cite{Yu2025DAFARL}, surveyed representative threats to smart inverters\cite{li2022cybersecurity}, and developed resilient control and mitigation strategies against different attack types\cite{ahn2023overview}. These efforts have significantly improved the understanding of cyber risks in converter-dominated systems. Nevertheless, most existing studies remain focused on attack modeling, detection, or mitigation for specific attack scenarios, rather than quantitatively assessing how feasible adversarial manipulations translate into node-level vulnerability from the attacker’s perspective.

A critical challenge is that existing strength metrics in power electronics systems are design-oriented rather than attack-oriented. Indicators such as gSCR-, MISCR-\cite{Dong2019gSCR}, and IMR\cite{zhu2024IMR}-type metrics are valuable for assessing grid strength and small-signal robustness under nominal operating conditions, but they are not intended to characterize cyber-vulnerability under privilege-constrained attacks. In particular, they do not quantify how accessible destabilizing actions are when an attacker can coordinately manipulate multiple cyber-accessible variables within bounded operating and stealth constraints.
As a result, a node that appears sufficiently strong under conventional design criteria may still be highly vulnerable from an adversarial viewpoint. This reveals a fundamental research gap between nominal strength assessment and cyber-vulnerability assessment.

To address this problem, this letter develops an impedance-based Attack Reachable Domain (ARD) framework for cyber-vulnerability assessment in PESs. The main idea is to map bounded cyber-accessible manipulations to eigenvalue migration through impedance reshaping, and then evaluate node vulnerability through the geometry of the corresponding attack reachable domain. 
Based on this formulation, an Attack Penetration Index (API) is introduced as a continuous attacker-oriented metric that captures both the remaining distance to instability and the accessibility of successful destabilizing attacks within a privilege-constrained action space. In addition, to make the proposed assessment applicable when detailed inverter models are unavailable from an adversarial perspective, a practical gray-box implementation is established by integrating existing impedance identification and differentiable surrogate tools into the overall workflow. 
The main contributions of this letter are summarized as follows:

\begin{enumerate}
    \item This letter presents a novel attacker-oriented assessment paradigm for cyber-vulnerability analysis in power electronics systems.
    By quantifying how bounded cyber-accessible manipulations translate into eigenvalue drift through impedance reshaping, the proposed ARD framework enables node-level vulnerability ranking from the attacker’s perspective and provides actionable guidance for designers on control-access exposure and defense prioritization.
    \item An Attack Penetration Index (API) is developed as a unified attacker-oriented metric that captures both stability-margin erosion and the accessibility of attack-reachable instability, thereby enabling normalized node-level cyber-vulnerability comparison under privilege-constrained action spaces.
    \item A practical gray-box workflow is established by integrating existing impedance identification and differentiable surrogate tools, and the case studies on the modified test systems show that grid-strength indicators cannot substitute for adversarial vulnerability assessment.
\end{enumerate}

\vspace{-15pt}
\section{Problem Formulation}
\vspace{-3pt}
\subsection{Threat Model}

This letter considers a sophisticated attacker targeting grid-connected inverter-based resources (IBRs). Following the MITRE ATT\&CK for ICS framework \cite{Alexander2020ATTCKforICS}, the attacker reaches the inverter control layer through reconnaissance, initial access, and lateral movement, exploiting multiple entry points such as communication protocols, web-based human-machine interfaces, firmware update channels, and software supply chain vulnerabilities. Under a relatively realistic gray-box assumption, the attacker can perturb variables within the feasible attack set without access to internal inverter models or closed-form impedance expressions, consistent with practical adversarial and defender-side assessment settings.

Two representative classes of adversarial actions are considered in this work, corresponding to two different layers of the inverter control hierarchy:

\subsubsection{Operating Point Manipulation}
the attacker alters dispatch-level commands, including active/reactive power references and voltage magnitude reference. Although these variables do not directly modify the control law, they shift the steady-state operating condition and thereby reshape the inverter impedance indirectly.

\subsubsection{Control-parameter tampering}
the attacker directly modifies internal controller parameters, such as current/voltage loop bandwidths, virtual impedance gains, and Phase-Locked Loop(PLL) parameters. Compared with operating-point manipulation, these attacks reshape the inverter frequency-domain behavior more directly.

The overall adversarial action can be expressed as
\begin{equation}
v_{\mathrm{atk}}=\{x_{\mathrm{op,atk}},\rho_{\mathrm{atk}}\}\in\Omega
\label{eq:vatk}
\end{equation}
where $x_{\mathrm{op,atk}}$ denotes the manipulated operating point, $\rho_{\mathrm{atk}}$ denotes the tampered control parameters, and $\Omega$ is the feasible attack set determined jointly by protocol-valid command ranges, physical protection limits, accessible attack privileges, and stealth requirements. 


\vspace{-10pt}
\subsection{Attack Reachable Domain Formulation}
We characterize the reachable degradation of the critical stability mode of a power electronics system under privilege-constrained cyber manipulation. To this end, this letter defines the ARD as the set of all critical-eigenvalue locations reachable from the nominal operating condition under feasible adversarial actions.

Let $\lambda_0$ denote the baseline critical eigenvalue, for a given attack vector $v_{\mathrm{atk}}\in\Omega$, the induced impedance perturbation $\Delta Z$ is represented by
\begin{equation}
\Delta Z = F_Z(v_{\mathrm{atk}})
\label{eq:dZ}
\end{equation}
where $F_Z(\cdot)$ denotes the parameter-to-impedance mapping. Using the eigenvalue sensitivity $g_\lambda$ with respect to impedance perturbation, the corresponding eigenvalue drift $\Delta \lambda$ is written as
\begin{equation}
\Delta \lambda = g_\lambda \Delta Z = g_\lambda F_Z(v_{\mathrm{atk}}).
\label{eq:dlambda}
\end{equation}
Accordingly, the ARD $R_{\lambda}$ is defined as
\begin{equation}
R_\lambda=\left\{\lambda_0+\Delta\lambda \;\middle|\; \Delta\lambda=g_\lambda F_Z(v_{\mathrm{atk}}),\; v_{\mathrm{atk}}\in\Omega \right\}\subset\mathbb{C}.
\label{eq:ARD}
\end{equation}


The ARD provides a geometric characterization of attacker-oriented vulnerability. In this sense, the ARD captures not merely whether a system is nominally strong, but how effectively feasible cyber manipulations can penetrate its small-signal stability margin.

\begin{figure}[htbp]
\centering
\includegraphics[width=0.85\linewidth]{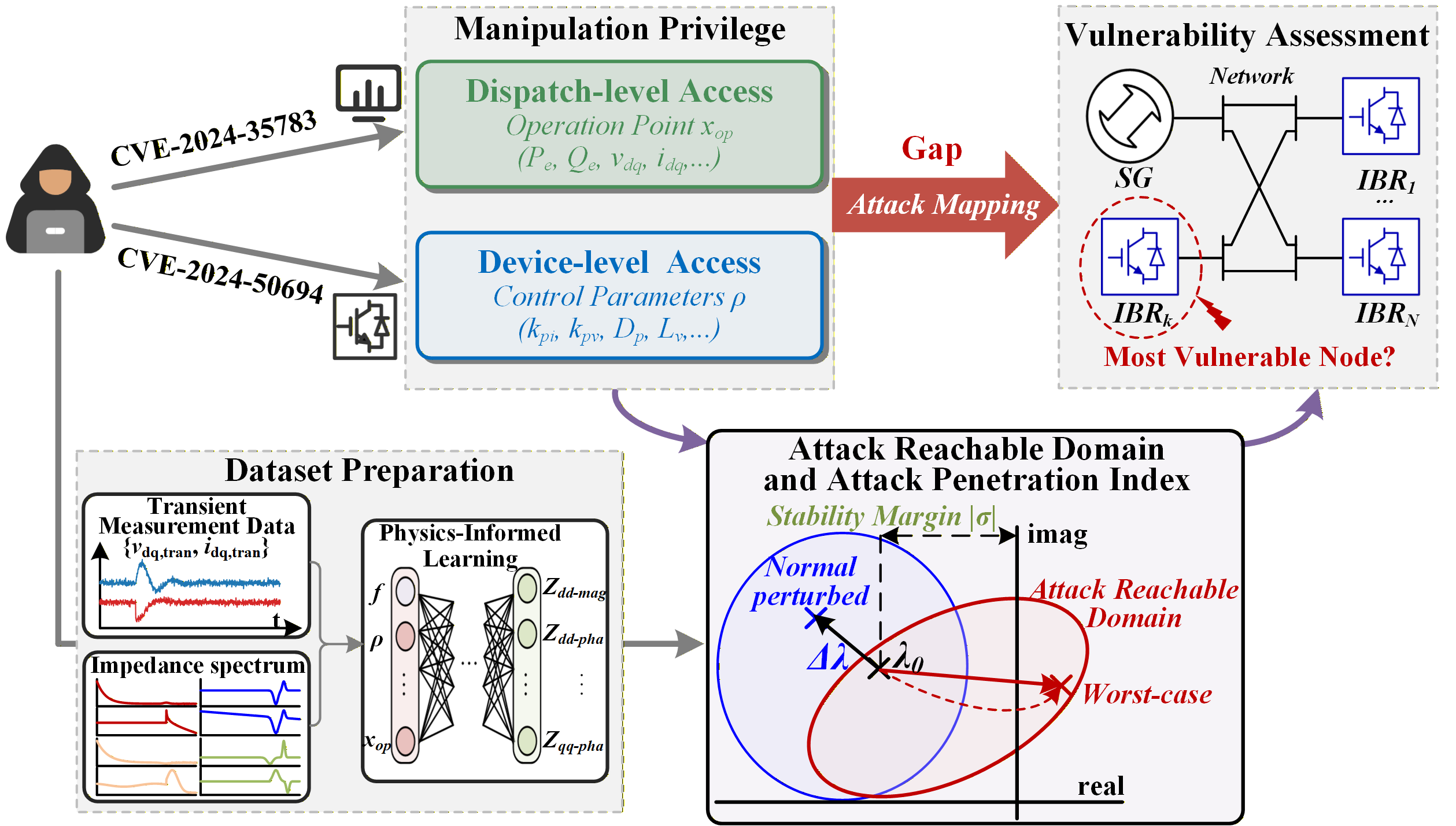}
\vspace{-1em}
\caption{Overview of the framework.}
\label{fig:Framework}
\end{figure}
\vspace{-14pt}

\section{The Proposed Metric}
Under gray-box conditions, neither the parameter-to-impedance mapping ${F}_{Z}(\cdot)$ nor the eigenvalue sensitivity $g_{\lambda}$ is directly available. This section develops a practical pipeline for ARD and API evaluation and a boundary-probing method for efficient computation.

\vspace{-14pt}
\subsection{Approximation of the Parameter to Impedance Mapping}

The differentiable surrogate~$\hat{F}_Z$ is first constructed from measurement data through impedance identification\cite{Fan2021ERA} and physics-informed surrogate training\cite{Li2025PINN}.

\subsubsection{Impedance Identification}
To build the training dataset, Latin Hypercube Sampling generates $N$ parameter combinations $\{v_i=(x_{\mathrm{op},i},\,\rho_i)\}_{i=1}^{N}$ over the parameter range defined by industry standards and physical operating limits.
For each~$v_i$, the inverter is configured to the corresponding operating point and control settings, and a small transient perturbation is injected at the point of common coupling~(PCC).
The terminal voltage and current responses are processed by the Eigensystem Realization Algorithm~(ERA) to identify the sub-synchronous impedance:
\begin{equation}\label{eq:ERA}
  Z_{\mathrm{inv},i}(s)
  = \mathcal{F}_{\mathrm{ERA}}\!\!\left(
      v_{dq,\mathrm{tran}}^{(1,2)}(t),\;
      i_{dq,\mathrm{tran}}^{(1,2)}(t)
    \right)\bigg|_{v_i}
\end{equation}
yielding the labeled dataset 
\begin{equation}\label{eq:dataset}
  \mathcal{D}
  = \bigl\{\bigl(v_i,\;Z_{\mathrm{inv},i}\bigr)\bigr\}_{i=1}^{N}
\end{equation}
in which each sample pairs a feasible attack-related parameter vector with the corresponding identified impedance spectrum.
 
\subsubsection{Physics-Informed Surrogate Training}
The surrogate~$\hat{F}_Z$ is trained on~$\mathcal{D}$ to learn a continuous mapping from attack-related parameters to inverter impedance.
A useful physical prior is that the dq-frame impedance of an inverter can be expressed in a rational polynomial form:~\cite{Li2025PINN}
\begin{equation}\label{eq:Zmn}
  Z_{mn}(s)
  = \frac{\mathbf{x}^{T}\mathbf{A}_{mn}\,\mathbf{x}}
         {\mathbf{x}^{T}\mathbf{A}_{0}\,\mathbf{x}},
  \quad m,n\in\{d,q\}
\end{equation}
where $\mathbf{x}$ is a polynomial basis constructed from operating-point variables, and $\mathbf{A}_{mn}$,$\mathbf{A}_0$ are coupling matrices that depend on control parameters.
Accordingly, the surrogate comprises a physics-based branch that computes~$\mathbf{x}$ from~$x_{\mathrm{op}}$ through hard-coded algebraic relationships, and a neural-network branch that learns the mapping $\rho$ to $\mathbf{A}_{mn},\mathbf{A}_0$.

\vspace{-13pt}
\subsection{Baseline Assessment and Boundary-Oriented Optimization}
With the trained surrogate $\hat{F}_Z$, the eigenvalue sensitivity can now be assembled.
Under the baseline condition, the inverter impedance spectrum $Z_\mathrm{inv}(j\omega)$ is evaluated using the trained surrogate. Meanwhile, the grid-side equivalent impedance $Z_\mathrm{g}(j\omega)$, representing the Thevenin equivalent seen from the target bus, is obtained from the remaining network using phasor measurement unit (PMU) steady-state data. The system-level admittance matrix $\hat{Y}_\mathrm{sys}$ is then constructed as 
\begin{equation}\label{eq:system-imepedence}
    \hat{Y}_\mathrm{sys}(j\omega) = (Z_\mathrm{inv}(j\omega)+Z_\mathrm{g}(j\omega))^{-1}
\end{equation}

Applying vector fitting to $\hat{Y}_\mathrm{sys}(j\omega)$ yields the baseline critical eigenvalue $\lambda_0$. For an induced inverter-impedance perturbation $\Delta Z_{\mathrm{inv}}$, the corresponding eigenvalue drift satisfies~\cite{zhu2021participation}:
\begin{equation}\label{eq:dlambda-PZ}
    \Delta \lambda = \langle P_{\lambda}, \Delta Z_{\mathrm{inv}}(\lambda_0) \rangle
\end{equation}
where 
\begin{equation}\label{eq:Plambda}
    P_{\lambda}=-\mathrm{Res}_{\lambda_0}^{*}\hat{Y}_\mathrm{sys}
\end{equation}
is the impedance participation factor matrix, defined as the conjugate transpose of the residue matrix of $\hat{Y}_\mathrm{sys}$ evaluated at the pole $\lambda_0$. 
Since the system is represented in the dq frame, both $Z_{\mathrm{inv}}$and $P_\lambda$ are $2\times2$ matrices. The scalar $g_\lambda$ in \eqref{eq:dlambda} can therefore be regarded as the vectorized form of $P_\lambda$, while the matrix form in \eqref{eq:dlambda-PZ} is used hereafter for compactness and physical interpretability.

With $\hat{F}_Z$ and $P_\lambda$ established, the ARD defined in \eqref{eq:ARD} becomes computable for any $v_\mathrm{atk} \in \Omega$. 
To avoid exhaustive enumeration over a multi-dimensional feasible attack set, we further introduce a boundary-oriented optimization to identify the ARD boundary point:
\begin{equation}
\begin{aligned}
\max_{v_{\mathrm{atk}}}\quad & \mathrm{Re}(\Delta\lambda) \\
\text{s.t.}\quad
& v_{\mathrm{atk}}=\{x_{\mathrm{op,atk}},\rho_{\mathrm{atk}}\}\in\Omega, \\
& d_{\mathrm{BDD}}(x_{\mathrm{op,atk}})<\epsilon_1,\quad d_{\mathrm{IDS}}(\rho_{\mathrm{atk}})<\epsilon_2 .
\end{aligned}
\label{eq:opt}
\end{equation}
Here, $d_{\mathrm{BDD}}(\cdot)$ and $d_{\mathrm{IDS}}(\cdot)$ denote that operating point deviations remain below the bad data detection (BDD) threshold and control parameter perturbations evade intrusion detection systems (IDS), respectively.

By automatic differentiation through $\hat{F}_Z$, the gradient of the boundary-search objective can be written as
\begin{equation}\label{eq:gradient}
    \nabla_{v_{\text{atk}}} \text{Re}(\Delta\lambda) = \text{Re} \left(\langle P_{\lambda} ,\frac{\partial \hat{F}_Z(v_{\text{atk}})}{\partial v_{\text{atk}}} \rangle \right)
\end{equation}
and the iterate is updated as
\begin{equation}\label{eq:worst-case}
    v_{\text{atk}}^{(t+1)} = \Pi_{\Omega} \left( v_{\text{atk}}^{(t)} + \alpha \nabla_{v_{\text{atk}}} \text{Re}(\Delta\lambda) \right)
\end{equation}
where $\alpha$ is the step size and $\Pi_\Omega$ projects onto the feasible stealth-constrained set. Upon convergence, $v_{\text{atk}}^*$ yields the largest real-part drift $\Delta\lambda_{\max}$, which serves as a limiting sample for subsequent API evaluation.

\vspace{-10pt}
\subsection{Attack Penetration Index}
Based on the ARD, the API is defined as a normalized metric for node-level cyber-vulnerability comparison:

\begin{equation}
\mathrm{API} = \left\{
\begin{array}{ll}
\displaystyle
\frac{\max_{v_{atk}\in\Omega}\text{Re}(\Delta\lambda(v_{atk}))}{|\text{Re}(\lambda_{0})|}, 
& \text{if } \forall \text{Re}(\lambda(v_{atk}))<0 \\[6pt]
1 + \frac{\iint_{R_{\lambda} \cap \mathbb{C}^+} d\sigma d\omega}{\iint_{R_{\lambda}} d\sigma d\omega}, 
& \text{if } \exists \text{Re}(\lambda(v_{atk})) \ge 0 
\end{array}
\right.
\label{eq:API_definition}
\end{equation}
where $\lambda(v_{atk}) = \lambda_{0} + \Delta\lambda(v_{atk})$ and  $C^+$ denotes the unstable right-half complex plane.



The API provides a unified attacker-oriented measure of node-level cyber-vulnerability. When the ARD remains entirely in the stable region, it reduces to the maximum real-part drift normalized by the baseline damping $|\mathrm{Re}(\lambda_0)|$, characterizing stability-margin erosion. Once the ARD crosses the imaginary axis, it becomes the fraction of the reachable domain lying in the unstable region, characterizing the accessibility of destabilizing attacks. Thus, $\mathrm{API}=1$ marks the transition from margin erosion to attack-reachable instability.

For each target bus $i$, the API is first evaluated for each critical oscillatory mode 
$k$ with non-negligible participation from that bus. The bus-level index reported for vulnerability ranking is then defined as
\begin{equation}\label{eq:bus-api}
    \mathrm{API}_{i} = \max_{k \in \mathcal{K}_i} \mathrm{API}_{i,k}
\end{equation}
where $\mathcal{K}_i$ denotes the set of dominant modes associated with bus $i$. This aggregation is consistent with attacker-oriented assessment that the vulnerability of a node should be determined by the most attack-reachable destabilizing mode.

Unlike the grid-strength indicators such as gSCR, MISCR, and IMR, API is attacker-oriented rather than design-oriented, as it explicitly incorporates privilege constraints that define the feasible attack set, and it optimizes over the worst-case direction rather than assuming arbitrary or uniform perturbations. 

\vspace{-10pt}
\section{Case Studies}

Two case studies are conducted to validate the proposed ARD framework and the resulting API. The 4-bus case verifies cross-layer attack amplification and the physical interpretation of the API threshold, while the modified IEEE 39-bus case demonstrates node-level vulnerability ranking in a larger inverter-dominated system. Since impedance identification and physics-informed learning are proven techniques, the corresponding results are omitted here for brevity and available in~\cite{github}.
\vspace{-10pt}

\begin{figure}[htbp]
\centering
\includegraphics[width=0.8\linewidth]{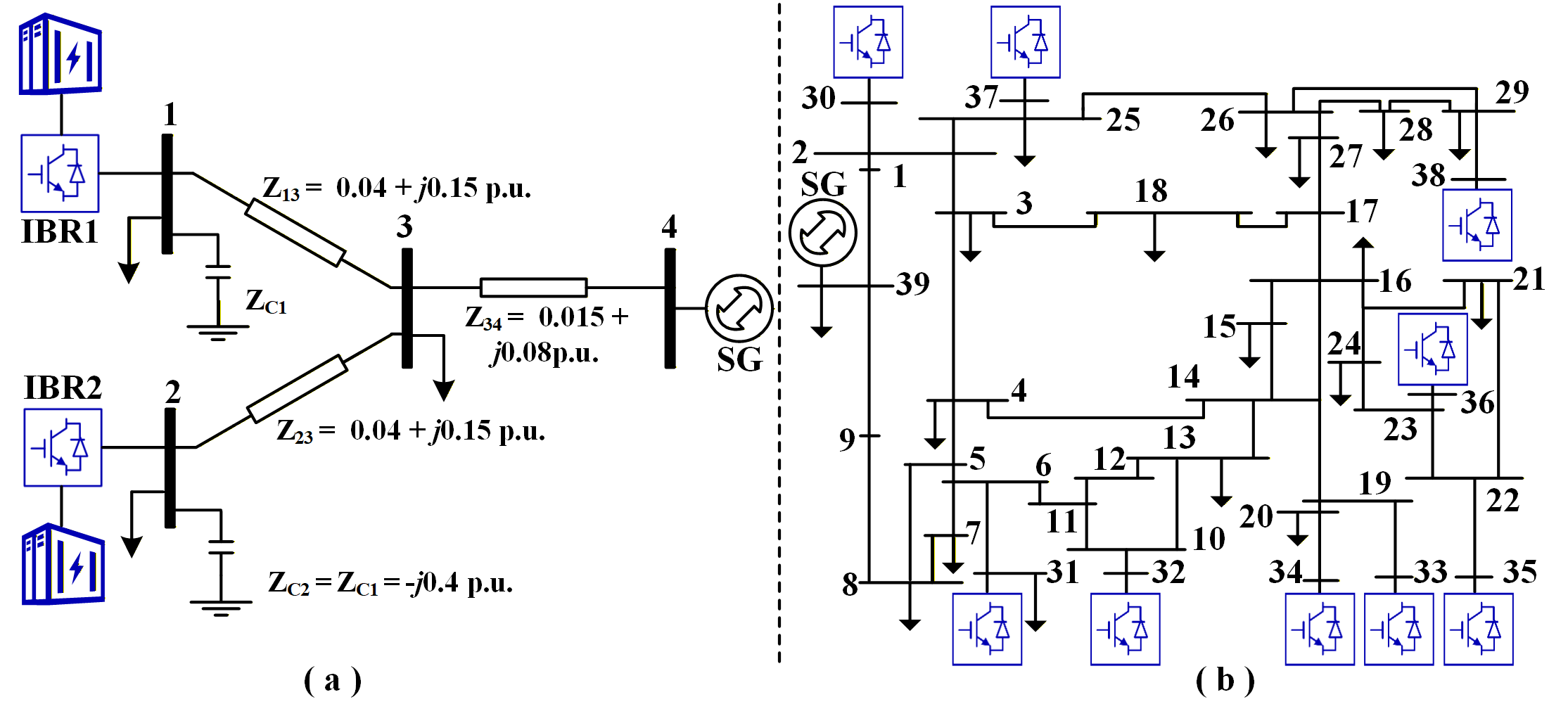}
\vspace{-1em}
\caption{Single-line diagram of the 4-bus system and modified IEEE-39Bus system. Detailed system parameters are available in \cite{github}.}
\label{fig:4BusSystem}
\end{figure}

\vspace{-16pt}
\subsection{4-Bus Test System}
A 4-bus benchmark system composed of one synchronous generator and two grid-forming virtual synchronous generator (VSG)-controlled IBRs is established, as shown in Fig.~\ref{fig:4BusSystem}(a). Two dominant oscillatory modes are identified around the nominal operating point.

\begin{figure}[htbp]
\centering
\includegraphics[width=0.85\linewidth]{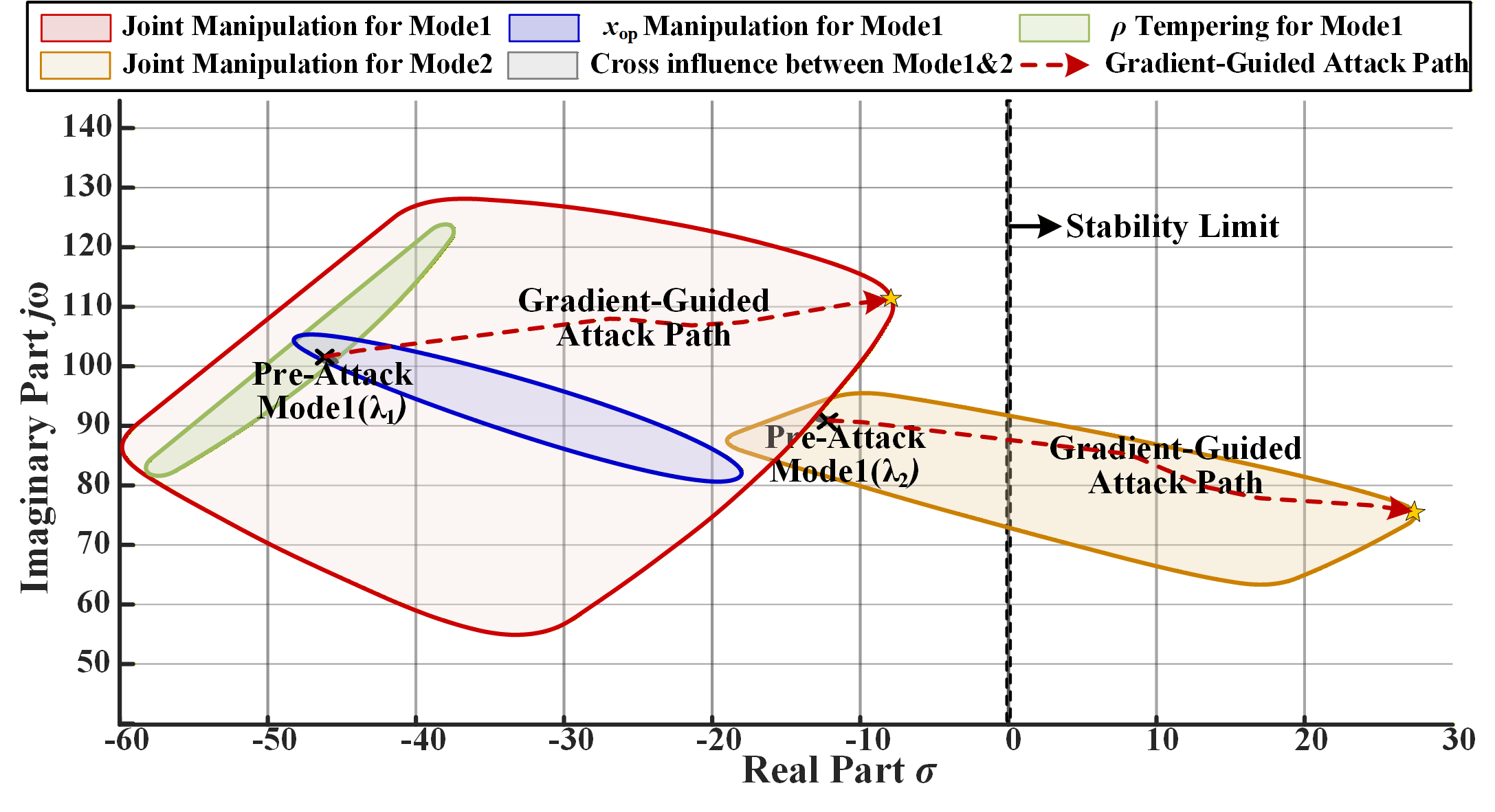}
\vspace{-1em}
\caption{Calculated ARDs of the 4-bus system for the two dominant modes under single-layer and joint cyber manipulations.}
\label{fig:ARR}
\end{figure}

Fig.~\ref{fig:ARR} presents the corresponding ARDs under operating-point manipulation, control-parameter tampering, and their coordinated joint manipulation, while Table~\ref{tab:API-4Bus} summarizes the associated API values. The values in parentheses denote the dominant oscillation frequency under the corresponding worst-case attack.

The results show that joint manipulation consistently yields the largest reachable domain and the highest API, revealing a pronounced cross-layer amplification effect. In particular, for Mode 2, the joint attack raises the API to 1.621, exceeding both single-layer cases and driving the ARD across the stability boundary. By contrast, Mode 1 remains entirely in the stable left-half plane with $\text{API}<1$, indicating that the most damaging feasible attack can only erode the damping margin without making instability attack-reachable. These two modes therefore illustrate the physical interpretation of the API threshold. Moreover, the gradient-guided search converges to the rightmost boundary point of each ARD, which is consistent with the boundary-oriented extreme-test generation in \eqref{eq:opt}.

\begin{figure}[htbp]
\centering
\includegraphics[width=0.9\linewidth]{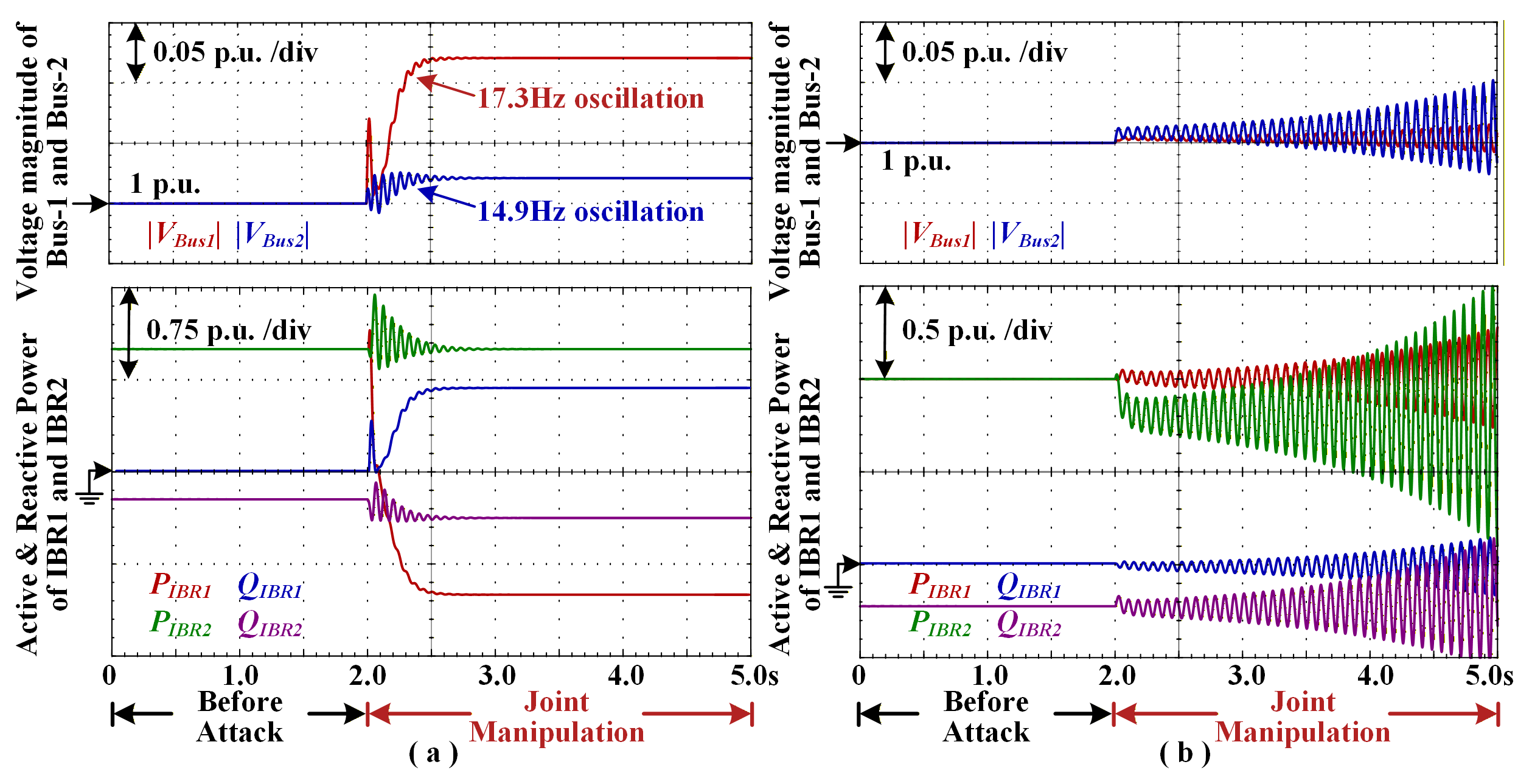}
\vspace{-1em}
\caption{Time-domain responses under the worst-case feasible attacks for (a) Mode 1; (b) Mode 2}
\label{fig:4BusSystemWaveform}
\end{figure}

The frequency-domain prediction is further supported by the time-domain responses in Fig.~\ref{fig:4BusSystemWaveform}. Under the worst-case attack associated with Mode 1, the system exhibits a bounded oscillatory transient whose dominant frequency agrees well with the ARD prediction. In contrast, the worst-case attack associated with Mode 2 leads to growing oscillations and eventual instability, consistent with $\mathrm{API} > 1$ and the ARD penetrating into the right-half plane.

\vspace{-13pt}
\begin{table}[htbp]
\centering
\footnotesize
\caption{API values for the 4-bus system}
\vspace{-1em}
\label{tab:API-4Bus}
\setlength{\tabcolsep}{3pt}
\renewcommand{\arraystretch}{1.15}
\begin{tabular}{c*{3}{c}}
\toprule
\textbf{Mode} & \textbf{$x_{op}$ Manipulation} & \textbf{$\rho$ Tampering} & \textbf{Joint} \\
\midrule
$\lambda_1$(16.4 Hz) & 0.644 (13.1 Hz) & 0.178 (19.7 Hz) & 0.889 (17.8 Hz) \\
$\lambda_2$(14.3 Hz) & 0.431 (11.6 Hz) & 0.807 (13.5 Hz) & 1.621 (\textbf{Unstable}) \\
\bottomrule
\end{tabular}
\end{table}

\vspace{-13pt}

\vspace{-18pt}
\subsection{Modified IEEE 39-Bus Test System}

To validate the scalability of the proposed metric, the IEEE 39-bus system is modified by replacing nine generators other than Bus 39 with IBRs, as shown in Fig.~\ref{fig:4BusSystem}(b). The resulting system remains nominally stable with a gSCR of 4.845, and all IBR buses satisfy nominal-strength requirements. This provides a suitable benchmark for comparing nominal electrical-strength indicators with the proposed attacker-oriented vulnerability metric. Table~\ref{tab:scr_imr} summarizes the MISCR, IMR, and maximum bus-level API values of all IBR buses.

\vspace{-13pt}
\begin{table}[h]
\centering
\footnotesize
\caption{Nominal strength indicators and API values of IBR buses}
\vspace{-1em}
\label{tab:scr_imr}
\setlength{\tabcolsep}{3pt}
\renewcommand{\arraystretch}{1}
\begin{tabular}{l*{9}{c}}
\toprule
\textbf{Bus} & \textbf{30} &\textbf{31} &\textbf{32} & \textbf{33} & \textbf{34}& \textbf{35} & \textbf{36} & \textbf{37} & \textbf{38}\\
\midrule
MISCR & 14.163 & 4.694 & \textbf{4.277} & 5.468 & 5.018 & 5.425 & \textbf{4.703} & \textbf{5.111} & 1.855 \\
IMR   & 0.873  & 0.213 & \textbf{1.103} & 0.282 & 0.139 & 0.247 & \textbf{0.734} & \textbf{0.420} & 1.735 \\
API   & 0.707  & 0.443 & \textbf{0.136} & 1.147 & 1.322 & 0.933 & \textbf{1.223} & \textbf{1.463} & 0.368 \\
\bottomrule
\end{tabular}
\vspace{4pt}
\end{table}

\vspace{-13pt}
The results show that API is not monotonic with either MISCR or IMR, indicating that nominal electrical strength and attacker-oriented cyber-vulnerability capture different properties. Bus 37 has the highest API, followed by Buses 34 and 36, whereas Bus 32 is only weakly attack-reachable. This ordering is not implied by nominal-strength indicators alone: Bus 30 has the highest MISCR yet remains below the instability boundary, while Buses 34, 36, and 37 satisfy grid-strength requirements but still have API values above unity. Therefore, nominal-strength indicators alone cannot identify the attack-vulnerable locations within the system.

\begin{figure}[h]
\centering
\includegraphics[width=0.9\linewidth]{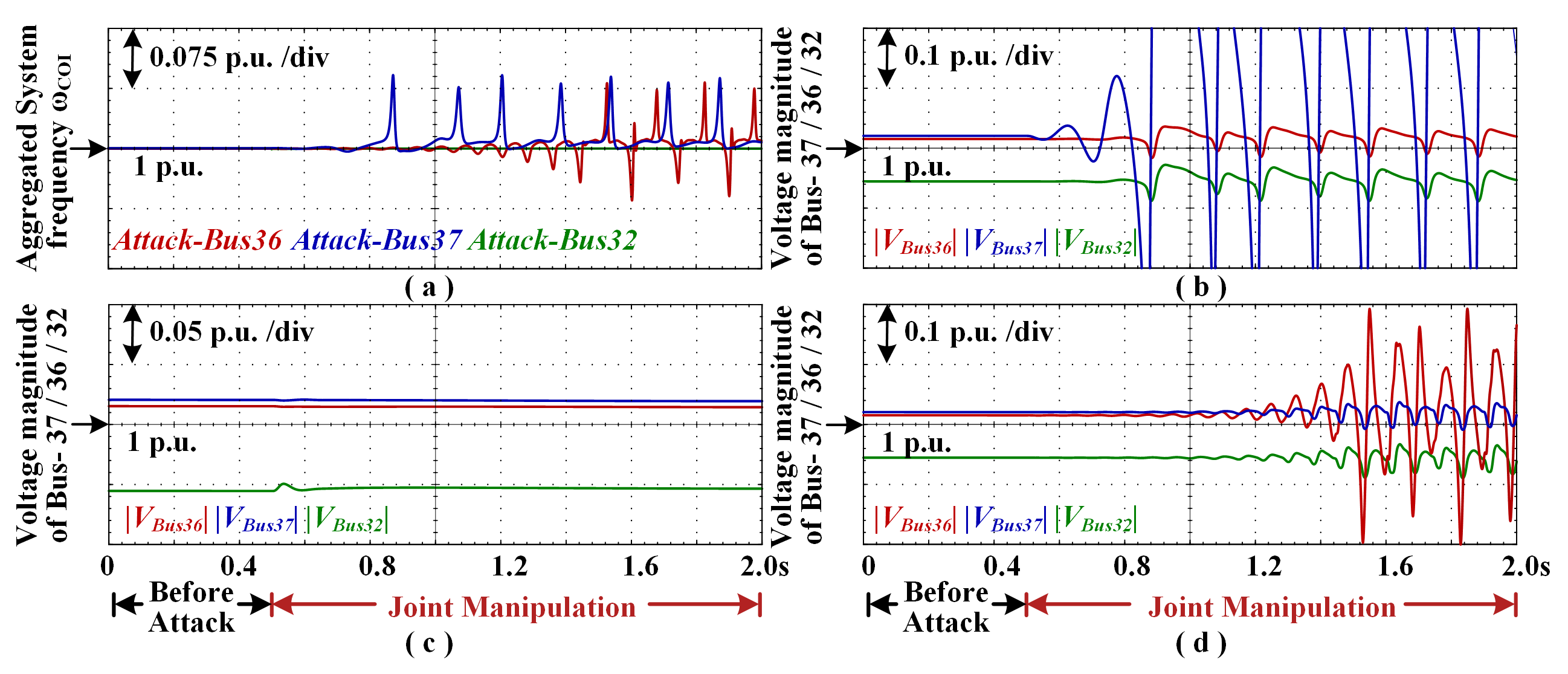}
\vspace{-1em}
\caption{Time-domain validation of bus-level vulnerability ranking when the worst-case joint manipulation targets Bus 37, 32, and 36, respectively. (a) Aggregated system frequency $\omega_\mathrm{COI}$; (b)-(d)Voltage-magnitude responses under the attack on Bus 37, 32, and 36, respectively.}
\label{fig:39BusSystemwaveform}
\end{figure}

To examine the practical meaning of the bus-level API ranking, Fig.~\ref{fig:39BusSystemwaveform} presents compact time-domain comparisons under the same attack privilege class, where $\omega_\mathrm{COI}$ is the inertia-weighted aggregated frequency of all VSG-controlled IBRs. The results show that worst-case joint attacks on higher-API buses induce more severe system-level disturbances. Attacks on Buses 37 and 36 lead to stronger oscillation growth and poorer recovery, whereas the response under attack on Bus 32 remains bounded and gradually decays. In particular, the pronounced vulnerability of Bus 36, overlooked by nominal-strength indicators, is correctly captured by the proposed API. These results confirm API as a more informative basis for vulnerability ranking and defense prioritization.

\vspace{-13pt}

\vspace{-16pt}
\section{Conclusion}

This letter develops an impedance-based Attack Reachable Domain framework for cyber-vulnerability assessment in power electronics systems and introduces the Attack Penetration Index for node-level vulnerability comparison under privilege-constrained attacks. Case studies on the 4-bus and modified IEEE 39-bus systems show that coordinated cross-layer manipulations are substantially more damaging than isolated single-layer attacks. The results further show that nominal strength indicators cannot substitute for adversarial vulnerability assessment and highlight the need to transition from single-parameter monitoring to impedance-based dynamic defense mechanisms to secure inverter-dominated power grids. Future work will extend the framework to address large-signal stability under coordinated multi-node attacks and validate the approach on hardware-in-the-loop platforms.

\vspace{-9pt}
\bibliographystyle{IEEEtran}
\bibliography{mybibfile}

\end{spacing}

\end{document}